\documentstyle[aps,tighten]{revtex}
\begin{document}
\draft
\newcommand{\Fstar}{\raisebox{.2ex}{$\stackrel{*}{F}$}{}}
\newcommand{\astar}{\raisebox{-0.1ex}{$\stackrel{*}{a}$}{}}
\renewcommand{\thefootnote}{\fnsymbol{footnote}}
\title{Electromagnetic wave propagation inside a material medium: an
effective geometry interpretation}
\author{%
V. A. De Lorenci 
and  M. A. Souza} 
\address{ 
Instituto de Ci\^encias - Escola Federal de Engenharia de Itajub\'a \\
Av. BPS 1303 Pinheirinho, 37500-903 Itajub\'a, MG -- Brazil \\
(Electronic mail: {\tt lorenci@cpd.efei.br})} 

\date{\today}
\maketitle

\begin{abstract}
\hfill{\small\bf Abstract}\hfill\smallskip
\par
We present a method developed to deal with
electromagnetic wave propagation inside a material medium that
reacts, in general, non-linearly to the field strength.
We work in the context of Maxwell\rq s theory in the low 
frequency limit and obtain a geometrical representation  of light
paths for each case presented. The isotropic case and artificial
birefringence caused by an external electric field are analyzed as 
an application of the formalism and the effective geometry
associated to the wave propagation is exhibited. 
\end{abstract}
\renewcommand{\thefootnote}{\arabic{footnote}}

\section{Introduction}

The characteristics of electromagnetic waves propagating in 
vacuum  are governed by a second order partial differential
equation. However,
when propagating through dispersive media, such differential
equation is no longer satisfied and, in general, it is necessary to
deal with a complicated system of equations. As a consequence,
several phenomena have to be analyzed closer, like polarization
states, wave velocity and its paths inside the particular medium
where it is traveling. Indeed, such deviation from the 
linear regime is also found to occur in vacuum
when the field strength goes beyond its critical value
($E_{c} \sim B_{c} = m^2c^2/e\hbar$), which is the limit of
applicability\cite{Schwinger} of the linear Maxwell\rq s theory.
The theoretical description of light propagation in nonlinear
electrodynamics was first presented by Bialynicka-Birula
and Bialynicki-Birula\cite{Birula} and also by Adler\cite{Adler},
where the probability of the photon splitting under a strong 
external electromagnetic field was studied.
Dittrich and  Gies\cite{Dittrich} obtained the light 
cone conditions for a class of homogeneous non trivial QED
vacua by using the rule of 
average over polarization states. More recently
De Lorenci, Klippert, Novello and Salim \cite{Lorenci}
studied the same problem within a different formalism, in which
general light cone conditions
for a class of theories constructed
with the two gauge invariants ($F,G$) 
of the Maxwell field was derived
without making use of average over polarization states. 
There, birefringence phenomena \cite{Birefringence} is also 
described. Further, 
the same authors \cite{Novello} analyzed the 
geometrical aspects of wave propagation in the context of 
nonlinear electrodynamics and found that the paths of light
can be described in terms of an effective geometry which
represents a modification of the Minkowski metric. It depends
on the dynamics of the background electromagnetic field as
\begin{eqnarray}
g_{\scriptscriptstyle \pm}^{\mu\nu} =\eta^{\mu\nu} 
- \frac{4}{L_F}\left[[\left(
L_{FF} + \Omega_{\scriptscriptstyle \pm} L_{FG}\right)
F^{\mu}\mbox{}_{\lambda}F^{\lambda\nu} 
+ \left(L_{FG} + \Omega_{\scriptscriptstyle \pm} L_{GG}\right)
F^{\mu}\mbox{}_{\lambda}\Fstar^{\lambda\nu}\right]
\label{eq1}
\end{eqnarray}
where $\Omega_{\scriptscriptstyle \pm}$ are certain coefficients
depending on the dynamics and field strength, and $L_X$ represents
the derivative of the Lagrangian with respect to the invariant
$X=(F,G)$. Indeed, it is shown that photons 
described by nonlinear electrodynamics propagate as null geodesics 
in such effective metric. (See also Ref. \cite{VisserM} where the
effective geometry interpretation is considered in the analysis of
the Scharnhorst effect.) Therefore,
applying the previous formalism for material media is a difficult task, since
for most cases we do not know an equivalent Lagrangian expression describing
the physical system. 
Inside such media, Maxwell\rq s equations are supplemented by
the constitutive law that relates the electromagnetic 
excitations $D,B$ and field strengths $E,H$  with the quantities
describing the electric and magnetic properties of each medium.
The constitutive equations are, in general, nonlinear and
are determined phenomenologically. Sch\"onberg\cite{Schonberg} and
later Obukhov and Hehl\cite{Obukhov} derived the equivalent 
spacetime metric for Maxwell\rq s theory on an arbitrary 
(1+3)-dimensional manifold imposing a linear constitutive law. 
Within the same approach, Obukhov, Fukui and Rubilar \cite{Rubilar} 
investigated the wave propagation in the Maxwell electrodynamics 
with the most general linear constitutive law and derived the 
associated Fresnel equation which determines the wave normals 
from the constitutive coefficients.

In this paper we analyze the electromagnetic wave propagation
in material media in the geometrical optics 
approximation. We present
the general field equations  in terms of some functions depending
on the properties of the medium without making 
any restriction to the
constitutive law. The geometrical representation of the light
paths for some special cases are examined and the effective
metric associated to each situation are exhibited.

In order to achieve more simplicity, in what follows we work
in Minkowski spacetime employing a Cartesian coordinate system.
The background metric will be represented by $\eta_{\mu\nu}$,
which is defined by $diag(+1,-1,-1,-1)$. We set the units $c=1$.

\section{The field equations}

We define the antisymmetric tensors $F_{\mu\nu}$ and 
$P_{\mu\nu}$ representing the total electromagnetic field.
They can be expressed in terms of the strengths ($E,H$) and 
the excitations ($D,B$) of the electric and magnetic fields as
\begin{eqnarray}
F_{\mu\nu} &=& V_{\mu}E_{\nu} - V_{\nu}E_{\mu} 
- \eta_{\mu\nu}{}^{\alpha\beta}V_{\alpha}B_{\beta}
\label{1}
\\
P_{\mu\nu} &=& V_{\mu}D_{\nu} - V_{\nu}D_{\mu} 
- \eta_{\mu\nu}{}^{\alpha\beta}V_{\alpha}H_{\beta}
\label{2}
\end{eqnarray}
where $V_\mu$ represents the velocity 4-vector of an
arbitrary observer, which in Galilean coordinates will
be given by $V_\mu = \delta_{\mu}^0$. The Levi-Civita tensor
introduced above is defined in such way that $\eta^{0123} = +1$.
Since the electric and magnetic tensor fields contain only spatial
components, we will denote the products as
$E^{\alpha}E_{\alpha}=-E^{2}$ and $H^{\alpha}H_{\alpha}=-H^{2}$.
In general the properties of the media 
are determined by the tensors 
$\varepsilon_{\alpha\beta}$ and $\mu_{\alpha\beta}$ which
relate the electromagnetic excitation and the field strength by
the generalized  constitutive laws,
\begin{eqnarray}
D_{\alpha} &=&  \varepsilon_{\alpha}{}^{\beta}(E^\mu,H^\mu)E_{\beta}
\label{8}
\\ 
B_{\alpha} &=& \mu_{\alpha}{}^{\beta}(E^\mu,H^\mu)H_{\beta}.
\label{9}
\end{eqnarray} 
In the absence of sources Maxwell\rq s 
theory can be summarized by the equations
\begin{eqnarray}
V^{\mu}D^{\nu}{}_{,\nu} - V^{\nu}D^{\mu}{}_{,\nu} 
- \eta^{\mu\nu\alpha\beta}V_{\alpha}H_{\beta,\nu}&=&0
\label{16}
\\
V^{\mu}B^{\nu}{}_{,\nu} - V^{\nu}B^{\mu}{}_{,\nu} 
+ \eta^{\mu\nu\alpha\beta}V_{\alpha}E_{\beta,\nu}&=&0
\label{18}
\end{eqnarray} 
which is equivalent to $P^{\mu\nu}{}_{,\nu} = 0$ and
$\Fstar^{\mu\nu}{}_{,\nu} = 0$, respectively.
Therefore, the electromagnetic excitation is related to the
field strength by means of the constitutive relations (\ref{8}) and
(\ref{9}), whose derivatives with respect to the coordinates
can be presented as
\begin{eqnarray}
D_{\alpha,\tau} &=& \varepsilon_{\alpha}{}^{\beta}E_{\beta,\tau} 
+ \frac{\partial \varepsilon_{\alpha}{}^{\beta}}{\partial
E_\mu}E_{\beta}E_{\mu,\tau} + \frac{\partial       
\varepsilon_{\alpha}{}^{\beta}}{\partial H_\mu}E_{\beta}H_{\mu,\tau}
\label{28} 
\\
B_{\alpha,\tau} &=& \mu_{\alpha}{}^{\beta}H_{\beta_,\tau}
+\frac{\partial \mu_{\alpha}{}^{\beta}}{\partial E_\mu}
H_{\beta}E_{\mu,\tau} 
+\frac{\partial \mu_{\alpha}{}^{\beta}}{\partial
H_\mu}H_{\beta}H_{\mu,\tau}.
\label{29} 
\end{eqnarray}
Equations (\ref{16}) and (\ref{18}), together with the relations
(\ref{28}) and (\ref{29}),  are 
given in terms of the first derivatives of the electric and 
magnetic fields. They represent the
electromagnetic field equations inside an arbitrary material medium
whose properties are determined by the tensors 
$\varepsilon_{\alpha\beta}$ and $\mu_{\alpha\beta}$.

\section{Propagating waves}

In order to derive the relations that determine the propagating
waves, we will consider the Hadamard method of 
field discontinuities \cite{Lorenci,Hadamard}. 
Let us consider a surface of discontinuity $\Sigma$ defined by
$z(x^{\mu}) = 0$. Whenever $\Sigma$ is a global surface, it 
divides the spacetime  in two distinct regions $U^-$, for $z<0$,
and $U^+$, for $z>0$.  Given an arbitrary function of the
coordinates, $f(x^\alpha)$,  we define its discontinuity on 
$\Sigma$ as
\begin{equation}
\label{discontinuity}
\left[f(x^{\alpha})\right]_{\Sigma} 
\doteq \lim_{\{P^\pm\}\rightarrow P}
\left[f(P^+) - f(P^-)\right]
\end{equation}
where $P^+,\,P^-$ and $P$ belong to $U^+,\,U^-$ 
and $\Sigma$ respectively.
Applying these conditions 
for the electric and magnetic fields and their derivatives we set
\begin{eqnarray}
\left[E_{\mu}\right]_{\Sigma} = 0; \;\;\;
\left[E_{\mu ,\nu}\right]_{\Sigma} &=& e_{\mu}K_{\nu}
\label{34}
\\
\left[H_{\mu}\right]_{\Sigma} = 0; \;\;\;
\left[H_{\mu ,\nu}\right]_{\Sigma} &=& h_{\mu}K_{\nu}
\label{35}
\end{eqnarray}
where $e_{\mu}$ and $h_{\mu}$ represent the discontinuities of 
the fields on the surface $\Sigma$ and $K_{\lambda}$ is the wave 
propagation 4-vector.
Applying these conditions to the field equations (\ref{16}) and 
(\ref{18}), we obtain the
following set of equations governing the wave propagation
\begin{eqnarray}
\varepsilon^{\alpha\beta}K_{\alpha}e_{\beta} 
+\frac{\partial \varepsilon^{\alpha\beta}}{\partial   
E_\mu}K_{\alpha}E_{\beta}e_{\mu} +\frac{\partial
\varepsilon^{\alpha\beta}}{\partial H_\mu}
K_{\alpha}E_{\beta}h_{\mu} &=& 0
\label{41} 
\\ 
\mu^{\alpha\beta}K_{\alpha}h_{\beta} 
+\frac{\partial \mu^{\alpha\beta}}{\partial E_\mu}
K_{\alpha}H_{\beta}e_{\mu} +\frac{\partial
\mu^{\alpha\beta}}{\partial H_\mu}
K_{\alpha}H_{\beta}h_{\mu} &=& 0
\label{42} 
\\
\left(\varepsilon^{\mu\beta}e_{\beta} 
+\frac{\partial \varepsilon^{\mu\beta}}{\partial   
E_\alpha}E_{\beta}e_{\alpha} +\frac{\partial     
\varepsilon^{\mu\beta}}{\partial   
H_\alpha}E_{\beta}h_{\alpha}\right)(KV) +
\eta^{\mu\nu\alpha\beta}K_{\nu}V_{\alpha}h_{\beta} &=& 0  
\label{43} 
\\
\left(\mu^{\mu\beta}h_{\beta} 
+\frac{\partial \mu^{\mu\beta}}{\partial  
E_\alpha}H_{\beta}e_{\alpha} 
+\frac{\partial      
\mu^{\mu\beta}}{\partial H_\alpha}H_{\beta}h_{\alpha}
\right)(KV) -
\eta^{\mu\nu\alpha\beta}K_{\nu}V_{\alpha}e_{\beta} &=& 0
\label{44}
\end{eqnarray}
where we have defined $(KV) \doteq K^{\mu}V_{\mu}$.
Indeed, equations (\ref{41}) and (\ref{42}) follow from the
zeroth component of equations (\ref{16}) and (\ref{18}), 
respectively. 

We will restrict our analysis for those cases where 
$\varepsilon_{\alpha\beta} = \varepsilon_{\alpha\beta}(E^\mu,H^\nu)$
and $\mu_{\alpha\beta} = \mu\eta_{\alpha\beta} $. (From 
the symmetry of the field equations,
all results we obtain are equally true for the tensor
$\mu_{\alpha\beta}$). Thus, from equation
(\ref{42}) results the identity $h^{\beta}K_{\beta} = 0$, 
and equation (\ref{44}) gives the following relation between  
$h_\alpha$ and $e_\alpha$: 
\begin{equation}
h^{\mu} = 
\frac{1}{\mu (KV)}\eta^{\mu\nu\alpha\beta}K_{\nu}V_{\alpha}
e_{\beta}.
\label{49}
\end{equation}
By introducing these results in equation (\ref{43}) it follows that
\begin{equation}
Z^{\mu\beta}e_{\beta} = 0
\label{53}
\end{equation}
with 
\begin{eqnarray}
Z^{\mu\beta}=
\mu (KV)^2\!\!\left(\!\varepsilon^{\mu\beta}
+\frac{\partial \varepsilon^{\mu\alpha}}{\partial   
E_\beta}E_{\alpha}\!\!\right)
+(KV)\frac{\partial      
\varepsilon^{\mu\rho}}{\partial H_\alpha}\eta_{\alpha}
{}^{\tau\sigma\beta}
E_{\rho}K_{\tau}V_{\sigma}
-K^{\mu}K^{\beta}+(KV)V^{\mu}K^{\beta}
+K^{2}\eta^{\mu\beta}-(KV)^{2}\eta^{\mu\beta}.
\label{54} 
\end{eqnarray}
Non-trivial solutions of equation (\ref{53}) can be found for 
such cases where $\det\left| Z^{\mu\beta} \right| = 0$.
This condition is known in the literature as the Fresnel equation.
Despite of its importance, we will not be interested in solving it
explicitly, but in analyzing some special cases for which an effective
metric structure associated to the wave propagation can be
derived. For the cases where the constitutive
laws are linear, the Fresnel equation was solved in
Ref. \cite{Obukhov,Rubilar}.

\section{The effective geometry for media with isotropic permittivity}

The special case of media with isotropic permittivity can be stated by
assuming $\varepsilon^{\mu\beta} = \epsilon(E,H)(\delta^{\mu\beta}
-V^{\mu}V^{\beta})$.
Introducing this condition in equation (\ref{53}), we obtain
\begin{eqnarray}
&&\left\{\!
\left[K^2\! -\! (1\!-\!\mu\epsilon)(KV)^2\right]\eta^{\mu\beta}\!
-\!(KV)^2 \frac{\mu}{E}\frac{\partial \epsilon}{\partial E} 
E^\mu E^\beta\! 
-\!(KV)\frac{1}{H}\frac{\partial
\epsilon}{\partial H}
\eta^{\tau\nu\alpha\beta}H_{\tau}K_{\nu}V_{\alpha}E^{\mu}
\!-\!K^{\mu}K^{\beta} + (KV)V^{\mu}K^{\beta} \!\right\}e_{\beta}\!
=\! 0. 
\label{e-5} 
\end{eqnarray}
This equation can be solved by expanding
$e_{\beta}$ as a linear combination of four linearly independent vectors, which
can be chosen to be $E_{\beta}$, $H_{\beta}$, $K_{\beta}$
and $V_{\beta}$ as 
$
e_{\beta} = \alpha_{\scriptscriptstyle 0} E_{\beta} +  
\beta_{\scriptscriptstyle 0} H_{\beta} + 
\gamma_{\scriptscriptstyle 0} K_{\beta} + 
\delta_{\scriptscriptstyle 0} V_{\beta}$. 
Thus, substituting such representation of 
$e_{\beta}$ in equation (\ref{e-5}) and considering the linear 
independence of the vectors used to expand it,  
it yields the following equations 
\begin{eqnarray}
&&\alpha_{\scriptscriptstyle 0}\left[ 
\frac{K^2}{(KV)^2} -1 + \mu\frac{\partial(\epsilon E)}{\partial E}
-\frac{1}{(KV)H}\frac{\partial
\epsilon}{\partial H}  
\eta^{\tau\nu\alpha\beta}H_{\tau}K_{\nu}V_{\alpha}E_{\beta} 
\right] 
- \beta_{\scriptscriptstyle 0} \left[ \frac{\mu}{E}
\frac{\partial \epsilon}{\partial E}
E^{\alpha}H_{\alpha} \right]  
-\gamma_{\scriptscriptstyle 0} \left[ \frac{\mu}{E}
\frac{\partial \epsilon}{\partial E}
E^{\alpha}K_{\alpha} \right] = 0  
\label{f-3} 
\\
&&\alpha_{\scriptscriptstyle 0} E^{\mu}K_{\mu} +
\beta_{\scriptscriptstyle 0} H^{\mu}K_{\mu} 
+\gamma_{\scriptscriptstyle 0} (1-\mu\epsilon)(KV)^2 
+ \delta_{\scriptscriptstyle 0} (KV) =0
\label{f-4} 
\\
&&\alpha_{\scriptscriptstyle 0} (KV)E^{\mu}K_{\mu} +
\beta_{\scriptscriptstyle 0} (KV)H^{\mu}K_{\mu} 
+\gamma_{\scriptscriptstyle 0}(KV)K^2 + 
\delta_{\scriptscriptstyle 0}
\left[K^2 + \mu\epsilon(KV)^2 \right]=0
\label{f-5} 
\\
&&\beta_{\scriptscriptstyle 0}
\left[K^2 -(1-\mu\epsilon)(KV)^2\right]=0 .
\label{f-6} 
\end{eqnarray}
The solution of this system results in the light cone condition
\begin{equation}
K^2 = (KV)^2\left[1 - \mu\frac{\partial(\epsilon E)}{\partial E}
\right] 
+ \frac{1}{\epsilon E}\frac{\partial \epsilon}{\partial E} 
E^{\alpha}E^{\beta}K_{\alpha}K_{\beta}
+(KV)\frac{1}{H}\frac{\partial \epsilon}{\partial H}
\eta^{\tau\nu\alpha\beta}H_{\tau}K_{\nu}V_{\alpha}E_{\beta}.
\label{g-5}
\end{equation}
The wave propagation set by equation ({\ref{g-5}) can be
presented as $g^{\mu\nu}K_{\mu}K_{\nu}=0$, where we have defined
the effective geometry
\begin{equation}
g^{\mu\nu} = \eta^{\mu\nu} - 
\left[1 - \mu\frac{\partial(\epsilon E)}
{\partial E}\right]V^{\mu}V^{\nu} -
\frac{1}{\epsilon E}\frac{\partial \epsilon}{\partial E}
E^{\mu}E^{\nu}  -  \frac{1}{2H}\frac{\partial \epsilon}{\partial H}
\eta^{\tau\alpha\beta ( \mu}V^{\nu )}
H_{\tau}V_{\alpha}E_{\beta}
\label{h-2}
\end{equation}
in which the notation ${\scriptstyle (\mu\nu )}$ 
designates the operation of symmetrization over the index, e.g.
$a^{(\mu\nu )} = a^{\mu\nu} + a^{\nu\mu}$.
In this way we see that $K_{\mu}$ is a null vector in the
effective geometry $g_{\mu\nu}$ which constitutes a 
deviation from Minkowski background due to the dielectric
media in which the wave propagates.
Indeed, it can be shown \cite{Novello} 
that $K_{\mu}$ satisfies the geodesic equation
$K_{\mu ;\lambda}K^\lambda = 0,$ 
where the semicolon designates the covariant derivative in the 
geometry determined by $g_{\mu\nu}$. Thus, since $K_\mu$ is a
null vector in such effective geometry, the integral curves of
$K_\mu$ will be null geodesics.

The phase velocity $v$ of the waves can be derived from
the light cone condition (\ref{g-5}) by imposing  
$K^2 = \omega^2 - |\vec{K}|^2$ and defining
$v = \omega/|\vec{K}|$, resulting
\begin{equation}
v^2 = \left[ \mu\frac{\partial (\epsilon E)}{\partial E} -
\frac{1}{\omega H}\frac{\partial \epsilon}{\partial H}
\vec{K}\cdot\vec{E}\times\vec{H}\right]^{-1}\left[  1+
\frac{1}{\epsilon E}\frac{\partial \epsilon}{\partial E}  
(\vec{E}\cdot \hat{k})^2\right] 
\label{isove}
\end{equation}
where $\hat{k}$ is an unit vector in the $\vec{K}$ direction:  
$\hat{k}=\vec{K}/|\vec{K}|$.

In the particular case of homogeneous and isotropic media,
where $\epsilon$ and $\mu$ do not depend on the field strength,
the effective geometry reduces to the Gordon \cite{Gordon}
metric $g^{\mu\nu} \equiv \eta^{\mu\nu}+
(\epsilon\mu-1)V^{\mu}V^{\nu}$. The light paths will
be determined by $ds^2=0$, where
$ds^{2}=(1/\epsilon\mu)dt^{2}-dx^{2}-dy^{2}-dz^{2}$. 
It follows that the phase velocity
associated to the electromagnetic wave is given, as expected, by
$v^2 = 1/{\epsilon\mu}$,
and the refraction index of the medium 
$n_{\scriptscriptstyle 0}=\sqrt{\epsilon \mu}$.
The same result can be obtained directly from
the general expression (\ref{isove}) by imposing $\epsilon$
constant.

\section{The effective geometry for artificial birefringence 
caused by an external electric field}

When an optically isotropic dielectric is submitted  to an
external electric field it is experimentally known that it
may become optically anisotropic -- the so called 
Kerr electro-optic
effect. Under such situation the dielectric behaves optically
like a uniaxial crystal with its optic axis parallel to the
field direction \cite{Landau}. Let us analyze this effect
using the  formalism developed before and derive the effective geometry
associated to the wave propagation.

We set the following relation between the dielectric tensor 
$\varepsilon_{\mu\nu}$ and the external field
\begin{equation}
\varepsilon^{\mu\beta} = \epsilon\,(\eta^{\mu\beta} - V^{\mu}V^{\beta})
-\alpha E^{\mu}E^{\beta}
\label{74}
\end{equation}
with $\epsilon$ and $\alpha$ constants.
For this case, taking the product of equation (\ref{53}) with
$E_\mu$, and by considering the relations (\ref{41}) and (\ref{49}),
we obtain the light cone condition: 
\begin{equation}
K^2 = \left[1-\mu\left(\epsilon + 3\alpha E^{2}\right)
\right](KV)^2 + \frac{ 2\alpha}{(\epsilon + \alpha E^{2})}
E^{\mu}E^{\nu}K_{\mu}K_{\nu}. 
\label{139} 
\end{equation}
Indeed, such condition can be expressed as
$g^{\mu\nu}K_{\mu}K_{\nu} = 0$,
where we have introduced the effective geometry
\begin{equation}
g^{\mu\nu} = 
\eta^{\mu\nu} + \left[\mu\left(\epsilon + 3\alpha E^{2}\right) -1
\right]V^{\mu}V^{\nu} - \frac{ 2\alpha}{(\epsilon + \alpha E^{2})}
E^{\mu}E^{\nu}. 
\label{140} 
\end{equation}

The discontinuities of the electromagnetic field inside a dielectric
that reacts to the action of an external field as stated by equation
(\ref{74}), propagates along null geodesics of this effective
geometry. 

The phase velocity yields  
\begin{equation}
v^{2} = \frac{1}{\mu(\epsilon + 3\alpha E^{2})} 
\left[1 + \frac{ 2\alpha}{\epsilon + \alpha E^{2}}
(\vec{E} \cdot \hat{k})^2\right].  
\label{150}
\end{equation}
As before, $\hat{k}$ is an unit vector in the $\vec{K}$ direction.  
We shall recognize two limiting
cases from such expression. For
$\vec{E} \cdot \hat{k} = 0$, which leads to
$v^{2}_{\perp} = 1/\mu(\epsilon + 3\alpha E^{2})$;
and for 
$ \vec{E}\cdot\hat{k} = |\vec{E}| $, which leads to
$v^{2}_{\parallel} = 1/\mu(\epsilon + \alpha E^{2})$.

In order to identify the birefringence phenomenon we shall
examine the eigen-vector problem stated by equation (\ref{53}).
Assuming the initial condition $\vec{E}\cdot\vec{K}=0$ 
we obtain 
\begin{equation}
\left\{ \left[ \mu(KV)^{2}(\epsilon + \alpha E^{2}) + K^{2} 
- (KV)^{2}\right]
\eta^{\mu\nu}  - 2\alpha\mu (KV)^{2}E^{\mu}E^{\nu}\right\}
e_\nu = 0.
\label{164}
\end{equation}
The solutions for this eigen-vector problem can be
achieved by expanding $e_{\mu}$ as a linear combination of the
linearly independent  vectors  $E_{\mu}$, $H_{\mu}$, $K_{\mu}$ 
and $V_{\mu}$, resulting in the two light cone
conditions on the wave propagation,
\begin{eqnarray}
K^{2} - (KV)^{2}\left[ 1 -\mu (\epsilon + 3\alpha E^{2})\right] 
&=& 0
\label{169}
\\
K^{2} - (KV)^{2}\left[ 1 -\mu (\epsilon + \alpha E^{2})\right] 
&=& 0.
\label{167}
\end{eqnarray}
As we see, from equation (\ref{139}), the condition (\ref{169})
is already expected. Therefore, we also obtain the second 
solution (\ref{167}) that corresponds to another
mode of light propagation. Performing the
computation of the phase velocity associated to this mode
we obtain $v^{2} = 1/\mu(\epsilon + \alpha E^{2})$, which 
is the wave existing in the case where 
$\vec{E}\cdot\hat{k} = |\vec{E}|$ from equation (\ref{139}). 
Such mode survives for all directions of propagation, and 
there will be an additional mode depending on the angle from the
wave propagation to the direction of the external electric
field. This mode is governed by the effective geometry
introduced by (\ref{140}). Thus, birefringence occur when an
optically isotropic dielectric is submitted to an external
electric field.

The effective geometries that govern the electromagnetic
wave inside the medium will be given by (\ref{140}), for
the ray whose velocity depends on the direction of
the external field, and by
$
g^{\mu\nu}  = \eta^{\mu\nu} + \left[\mu(\epsilon 
+ \alpha E^{2}) - 1
\right]V^{\mu}V^{\nu},$  
for the ray whose velocity does not depend on the 
direction of the external field.

For the limiting case, where $\vec{E}\cdot \vec{K}=0$,
the phase velocities are
determined by $v^2_{\perp}$ and $v^2_{\parallel}$, 
as calculated before 
and the index of refraction yields $n^{2}_{\perp} = \mu(\epsilon +
3\alpha E^{2})$ and $n^{2}_{\parallel} 
= \mu(\epsilon + \alpha E^{2})$,
respectively. In most cases the term in $\alpha$ 
represents only a small
correction of the values of $\epsilon$. 
Thus for electromagnetic waves propagating
inside a material medium in a perpendicular direction to that
defined by the external electric field, the difference between
the index of refraction of the two existing rays yields
\begin{equation}
n_{\perp} - n_{\parallel} = 
n_{\scriptscriptstyle 0}\left[\frac{\alpha E^2}{\epsilon}
- \frac{\alpha^2 E^4}{\epsilon^2} + {\cal O}(\alpha^3) \right]
\label{157}
\end{equation}
where $n_{\scriptscriptstyle 0} 
= (\epsilon\mu)^{1/2}$ is the index of refraction
of the isotropic case, as derived in the previous section. Finally
when $\alpha =0$ we recover the isotropic case. 

\section{Conclusion}

In this paper we have dealt with electromagnetic wave 
propagation inside a material medium. The general field
equations were presented in terms of the field strengths
$(E^\mu,H^\mu)$ and the tensors $\varepsilon_{\mu\nu}$
and $\mu_{\mu\nu}$ that characterize 
the properties of each medium
where the propagation occurs. By using the Hadamard method
we derived the set of equations that govern the wave 
propagation. We examined some special cases where the light
cone conditions can be obtained without solving the general
Fresnel equation and for these cases we presented the
associated effective geometry. As an application of the
formalism, we analyzed the isotropic media and the 
artificial birefringence caused by an
external electric field applied over 
a dielectric medium that reacts
nonlinearly to the field excitation. For this
case we obtained as a first order of approximation the
known result that the difference between the index of
refraction of the two limiting waves propagating in such
dielectric is proportional to the square of the external
field. 

The method developed here can by applied to study the 
electromagnetic wave propagation inside material media.
(We do not considered the frequency dependent
material properties.)
To do this we must know the expression for the tensors
$\varepsilon_{\mu\nu}$ and $\mu_{\mu\nu}$ in terms of both
the specific medium and the external field. 

An effective geometry can be derived for each situation
and can be used in the study of the properties
of light propagation. With such geometrical
description, we present tools for testing kinematic
aspects of gravitation in laboratory, an issue very much
addressed these days (see Refs. \cite{Visser,Leon,Mat,Barbati}).
For instance, we could ask about the possibility of 
formation of structures like event horizons, bending
of light and others, in laboratory.
 
A possible continuation of this work would be the comparison 
between the Lagrangian method resumed in the effective 
geometry stated by (\ref{eq1}) and that obtained for each
case that can be derived for non-linear media. 
Also deserves further investigation the general effective
geometry associated with an arbitrary material medium in terms
of the tensors $\varepsilon_{\mu\nu}$ and $\mu_{\mu\nu}$.

\acknowledgements
The authors are grateful to R. Klippert and N. Figueiredo for reading 
the manuscript.
This work was partially supported by {\em Conselho Nacional
de Desenvolvimento Cient\'{\i}fico e Tecnol\'ogico} (CNPq)
and {\em Funda\c{c}\~ao de Amparo \`a 
Pesquisa no Estado de Minas Gerais} (FAPEMIG) 
of Brazil.

\end{document}